\documentclass[aps,prb,twocolumn]{revtex4}
\usepackage{graphics,amsmath}
\begin{document}


\hyphenation{super-lat-tice semi-con-ductor}
\hyphenation{Ins-ti-tu-to Cien-cia Ma-te-ria-les Ma-drid}
\hyphenation{Con-se-jo Su-pe-rior In-ves-ti-ga-cio-nes}
\hyphenation{De-par-ta-men-to}
\hyphenation{In-ter-mi-nis-te-rial}
\hyphenation{Uni-ver-si-dad}
\hyphenation{Cam-pus}
\hyphenation{Can-to-blan-co}
\tighten

\title{First order transition and phase separation in pyrochlores with
colossal-magnetoresistance}
\author{P. Velasco, J. Mira*, F. Guinea, J. Rivas*, M.J. Mart\'{i}nez-Lope, J.A.
Alonso and J.L. Mart\'{i}nez}
\address{Instituto de Ciencia de Materiales de Madrid, C.S.I.C., Cantoblanco, E-28049%
\\
Madrid, Spain. \\
Dep. F\'{i}sica Aplicada. Universidad de Santiago de Compostela. E-15782\\
Santiago de Compostela. Spain}
\date{Received XX October 2001}

\begin{abstract}
Tl$_{2}$Mn$_{2}$O$_{7}$ pyrochlores present colossal magnetoresistance (CMR)
around the long range ferromagnetic ordering temperature (T$_{C}$). The
character of this magnetic phase transition has been determined to be first
order, by purely magnetic methods, in contrast to the second order character
previously reported by Zhao {\it et al.} (Phys. Rev. Lett. {\bf 83}, 219
(1999)). The highest CMR effect, as in Tl$_{1.8}$Cd$_{0.2}$Mn$_{2}$O$_{7}$,
corresponds to a stronger first order character. This character implies a
second type of magnetic interaction, besides the direct superexchange
between the Mn$^{4+}$ ions, as well as a phase coexistence. A model is
proposed, with a complete Hamiltonian (including superexchange and an
indirect interaction), which reproduce the observed phenomenology.
\end{abstract}








\maketitle

\newpage

Colossal magnetoresistance (CMR) has been described for (Tl$_{2}$Mn$_{2}$O$%
_{7}$- related pyrochlores (up to 10$^{6}$\%), around the ferromagnetic
ordering temperature (T$_{C}$). The Tl$_{2}$Mn$_{2}$O$_{7}$ compound
(undoped system) contains only Mn$^{4+}$, so that colossal magnetoresistance
(CMR) is not related to the Jahn-Teller effect nor the Double Exchange
mechanism, associated with the mixed valence Mn$^{3+}$-Mn$^{4+}$ in
manganese perovskites. Initial data analyses\cite
{nature,Science1,Science2,Mishra} attributed the long range ferromagnetic
ordering to a direct superexchange interaction Mn$^{4+}$-O-Mn$^{4+}$. Later
on, the critical exponents were measured by Zhao {\it et al.}\cite{Zhao}
indicating a second order character of the magnetic transition. These
exponent values were very close to that predicted for a near-neighbor (n-n)
Heisenberg 3D system. Hence, the system was considered as a well known n-n
Heisenberg system and a simple and coherent picture was established.
Subsequently, the decrease of T$_{C}$ on hydrostatic pressure data\cite
{Presion1} in Tl$_{2}$Mn$_{2}$O$_{7}$, as well as the big difference in T$%
_{C}$ between the isomorphous systems Tl$_{2}$Mn$_{2}$O$_{7}$, In$_{2}$Mn$%
_{2}$O$_{7}$, Lu$_{2}$Mn$_{2}$O$_{7}$ and Y$_{2}$Mn$_{2}$O$_{7}$, T$%
_{C}\simeq $130 K, 129 K, 15 K and 16 K, respectively\cite{Shimakawa},
pointed out to a more complex system than assumed until now. Recently,
Nu\~{n}ez-Regueiro and Lacroix\cite{NL01} made a careful calculation, using
a perturbation expansion in the Mn-O hopping term, which reproduces either
the difference in T$_{C},$ depending on the different ions (Tl, In, Lu and
Y), as well as the low pressure dependence of T$_{C}$. In order to explain
the increase of T$_{C}$ at much higher hydrostatic pressure \cite
{NL01,Tissen-presion}, a new indirect interaction between the Mn ($t_{2g}$)
localized band mediated by the Tl($6s$)-O($2p$)-Mn($e_{g}$) correlated bands
is taken into account.

There were several reports on the Tl$_{2}$Mn$_{2}$O$_{7}$ family that, from
the theoretical point of view, proposed to explain the long range
ferromagnetic ordering as a more complex system than a simple classical n-n
Heisenberg in 3D\cite{Mishra,NL01,1,VG01,ML98}. The simplest interpretation
implies a classical Heisenberg interaction (second order phase transition)
and a magnetic and transport channels almost decoupled\cite{Science2,Zhao}.
In that sense the character of the magnetic transition is a very important
issue because only a simple n-n Heisenberg 3D system is compatible with a
second order character of the magnetic phase transition.

The possibility of considering the magnetic phase transition as a
first-order one was first explored theoretically by Bean and Rodbell\cite
{Beam}. For that purpose they considered a compressible material with an
exchange interaction strongly dependent upon the interatomic spacing. They
found that in such case, for an hypothetical second-order transition, the
expansion of the Gibbs free energy in terms of magnetization should have, at
the Curie temperature, a null cuadratic term and a positive quartic one.
Banerjee\cite{B64}detected the essential similarity between this result and
the Landau-Lifshitz criterion \cite{Landau} and condensed them into one that
provides a tool to distinguish first-order magnetic transitions from
second-order ones. It consists on the analysis of the sign of the quartic
term of the Gibbs free energy\cite{B64}, which is quite simple to obtain
graphically, simply observing the slope of isotherm plots of $H/M$ $vs.$ $%
M^{2}$. A positive or negative slope indicates the second- or first-order
character of the transition, respectively. It is worth mentioning that this
procedure allows the identification of the character of the transition by
purely magnetic methods, in a very effective way, as sucessfully proved by
Mira {\it et al.} in La$_{2/3}$(Ca,Sr)$_{1/3}$MnO$_{3}$ perovskites \cite
{Mira1,Mira2}.

The purpose of the present letter is mostly to study\ in detail the
character of the long range ferromagnetic transition, in different samples
of the Tl$_{2}$Mn$_{2}$O$_{7}$ family, with different ordering temperatures
and CMR effects, as a fundamental issue to understand the magnetic
interactions in this system.

Polycrystalline samples of Tl$_{2}$Mn$_{2}$O$_{7}$, Tl$_{1.8}$Cd$_{0.2}$Mn$%
_{2}$O$_{7}$ and Tl$_{2}$Mn$_{1.8}$Sb$_{0.2}$O$_{7}$ pyrochlores were
prepared under high pressure conditions, from stoichiometric amounts of the
corresponding oxides, Tl$_{2}$O$_{3}$, MnO$_{2}$, CdO, Sb$_{2}$O$_{3}$. All
the details of the sample preparation and structural characterization (X-ray
and neutron diffraction) are given elsewhere\cite
{Alonso-Tlpuro,AlonsoPRB,AlonsoAPL}.The magnetic susceptibility was measured
with a Superconducting Quantum Interference Device (SQUID) magnetometer from
Quamtum Design (San Diego, USA) in the range from 2 K to 300 K and magnetic
fields up to 5T. Transport and magnetotransport measurements were performed
by the four-points contact technique inside a Physical Properties
Measurement System (PPMS) cryostat also from Quamtum Design in the range
from 2 K to 350 K and magnetic fields up to 9T. The specific heat was
measured inside the same cryostat (PPMS) with a quasi-adiabatic heat pulse
relaxation method. The thermopower measurements were done using a standard $%
\Delta $T constant method, in a temperature range from 5K to 400K.

The dc-magnetization measurements, performed at a field of 100 Oe, show a T$%
_{C}$ of 130 K for the undoped material, whereas Cd-doping decreases T$_{C}$
($\simeq $ 110 K), and Sb-doping increases it (T$_{C}\simeq $ 190 K). As
already reported\cite{Alonso-Tlpuro,AlonsoPRB,AlonsoAPL}
the resistivity change strongly between the three compounds (i.e. at 300K, Tl%
$_{2}$Mn$_{2}$O$_{7}$ is 20 $\Omega $cm, Tl$_{2}$Mn$_{1.8}$Sb$_{0.2}$O$_{7}$
is 0.2 $\Omega $cm and Tl$_{1.8}$Cd$_{0.2}$Mn$_{2}$O$_{7}$ is 30 k$\Omega $%
cm). In all cases the ferromagnetic ordering is accompanied by a sudden drop
in resistivity, suggesting the onset to a metallic state. From the bulk
magnetization and the transport data we could associate a large value of T$%
_{C}$ with a more metallic character, and a low value of the CMR effect
(Sb-doped sample). On the other hand, Cd doping leads a strong
metal-insulator transition with a variation of 7 orders of magnitude of the
resistivity (at zero magnetic field), and almost the same ferromagnetic
ordering temperature (with respect to the pure compound) presenting a large
CMR effect up to 10$^{6}$ (for Tl$_{1.8}$Cd$_{0.2}$Mn$_{2}$O$_{7}$), under
an applied magnetic field of 9T.

The temperature dependence of the thermopower (S) is presented in Fig. 1.
The value of S is negative in all the temperature range for the three
samples, which suggests that the charge carriers are electrons(negative Hall
resistance). The value of S for Tl$_{2}$Mn$_{1.8}$Sb$_{0.2}$O$_{7}$ is
almost 20 times smaller than for the other two compounds. The large value of
S for Tl$_{2}$Mn$_{2}$O$_{7}$ and Tl$_{1.8}$Cd$_{0.2}$Mn$_{2}$O$_{7}$ is
consistent with the small carrier density observed in these compounds (0.005
e$^{-}$/u.c. and 0.0002 e$^{-}$/u.c., respectively\cite
{Alonso-Tlpuro,AlonsoAPL}). In all the cases a linear behavior is observed
far from T$_{C}$ (low and high temperature), but a sharp increase of S
around T$_{C}$ is noticeable in the three compounds. This is specially
strong in the case of Tl$_{1.8}$Cd$_{0.2}$Mn$_{2}$O$_{7}$ where a sharp peak
is observed around T$_{C}$, which implies a big change in the slope dS/dT,
which could be related to a sharp variation of the charge carriers density
around T$_{C}$.

The specific heat data for the three compounds was measured in a wide
temperature range around T$_{C}$. In order to remove the phononic component
of the specific heat, we calculate it with an Einstein model with 3
oscillators centered at three frequencies (120K, 250K and 575K). The
subtraction of the phononic component from the total (measured) specific
heat is presented in Fig. 2 as the magnetic specific heat around T$_{C}$ for
the three compounds. The data clearly show the magnetic transition although
they do not assess on its character as first or second order.

In order to apply the criterion for the study of the character of the
magnetic transition from pure magnetic methods, initial magnetization
isotherms were measured around the respective T$_{C}$'s. Data were taken
with a SQUID between 0 and 10 kOe. Before each run, samples were heated up
to 300 K (well above their T$_{C}$'s) and cooled to the measuring
temperature under zero field, in order to ensure perfect demagnetization of
the samples.

Fig. 3 shows the results for Tl$_{2}$Mn$_{2}$O$_{7}$. At 130 K ($\simeq $ T$%
_{C}$), the curves show a small negative slope, indicating, according to
Banerjee's criterion, a first-order character of the transition. This
negative slope, found at low fields, continues above this temperature (see
inset). This fact is probably causing the ''unusual characteristics'' found
by Zhao {\it et al.} in the analysis of the critical behavior of the system 
\cite{Zhao}. The same happens for the Cd-substituted pyrochlore (Fig. 4),
with a negative slope starting from 110 K ($\simeq $ T$_{C}$), and for the
Sb-substituted one (Fig. 5), where the negative slope appears at 190 K. From
the above experimental data we conclude that the magnetic phase transition
for the three compounds is first order. In the case of Tl$_{2}$Mn$_{2}$O$_{7}
$ the transition is weakly first order, but for Tl$_{1.8}$Cd$_{0.2}$Mn$_{2}$O%
$_{7}$\ and Tl$_{2}$Mn$_{1.8}$Sb$_{0.2}$O$_{7}$\ it is clearly first order.

We assume that the magnetic properties of the pyrochlores are determined by
the Mn$^{4+}$ ions, which interact with a dilute band of conduction
electrons. There is a ferromagnetic direct interaction between the spins of
the Mn ions, $J$, and a local Kondo-like coupling between the Mn spins and
the conduction electrons, $J^{\prime }$\cite{NL01,1,VG01,ML98,GGA00}. The
hamiltonian can be approximated as: ${\cal H}=\sum_{k,s}\epsilon
_{k}c_{k,s}^{\dag }c_{k,s}-J\sum_{ij}{\bf \vec{M}}_{i}{\bf \vec{M}}%
_{j}-J^{\prime }\sum_{i}c_{i,s}^{\dag }\vec{\sigma}_{ss^{\prime
}}c_{i,s^{\prime }}{\bf \vec{M}}_{i}$

\bigskip

and $\epsilon _{k}=\hbar ^{2}|{\bf \vec{k}}|^{2}/2m_{eff}$. The
magnetization of the Mn ion at site $i$ is denoted as ${\bf \vec{M}}_{i}$,
and ${\bf \vec{s}}_{i}=\sum_{s,s^{\prime }}c_{i,s}^{\dag }\vec{\sigma}%
_{ss^{\prime }}c_{i,s^{\prime }}$ is the polarization of the carriers at
unit cell $i$. This hamiltonian is characterized by three parameters with
dimension of energy, $J,J^{\prime }$ and the bandwidth, $W$, which can be
written in terms of $m_{eff}$ and the lattice constant, $a$, as $W=(\hbar
^{2}\pi ^{2})/(2m_{eff}a^{2})$. In addition, we have to specify the number
of conduction electrons per unit cell, $n$, or, alternatively, the position
of the chemical potential, $\mu $. In the following, we will assume that $n$
can change as function of temperature and applied field, while $\mu $ is
constant, as suggested in recent experiments \cite{Ietal00}.

The above hamiltonian can be analyzed by a variety of methods. It can be
shown, that, in the limit of a highly diluted conduction band\cite{ML98},
polarons will be formed in the paramagnetic phase. The same arguments can be
used to proof that the stability of isolated small polarons imply the
tendency towards phase separation in the presence of a finite carrier
density \cite{GGA00}. This result, in turn, implies that the magnetic
transition becomes first order. This transition arises from the feedback of
the carriers on the Mn spins\cite{GGA00}. The coupling of the spin of the
conduction electrons to the Mn ions induce an effective interaction between
the Mn spins, which goes as $-J^{\prime }{\bf \vec{M}}{\bf \vec{s}}({\bf 
\vec{M}})$. If ${\bf \vec{s}}({\bf \vec{M}})$ changes sufficiently fast,
this coupling leads to a negative quartic term in the free energy of the Mn
spins. The strength of this term increases as the carrier density is reduced
and, when the carrier density is small enough, it overcomes the usual
positive quartic term which arises from the entropy of the Mn ions, leading
to a first order phase transition\cite{B64}.

In the following, we use the mean field approach developed in\cite{GGA00} to
study the properties of a system described by the above hamiltonian in the
presence of a magnetic field, and at constant chemical potential. The
results are plotted in Fig. 6. They have been obtained with the parameters $%
J=W/15$ and $J^{\prime }=W/5$. If we take $W\approx 1$eV, the Curie
temperature in the absence of free carriers is given by $kT_{0}=Jz/3=(2W)/15%
\approx 160$K. The parameter $J^{\prime }\approx 0.2$eV is a reasonable
value for the Kondo coupling between the spin of a Mn ion and a conduction
electron localized in the same unit cell. In order to compare different
carrier densities, we present results for fixed chemical potentials at $\mu
=-0.09W$ and $\mu =-0.01W$. For these parameters, there is a first order
transition at $T=1.006T_{0}$ with an abrupt change of the Mn magnetization
to $M=0.003M_{sat}$ and at $T=1.2T_{0}$ and $M=0.32M_{sat}$ respectively. In
the first case, the transition is weakly first order. The negative slope of $%
H/M$ $vs.$ $M^{2}$ is more pronounced when the transition is more strongly
first order, as expected. At low carrier density, the carriers become fully
polarized at lower magnetizations. The contribution of the polarization of
the carriers to the value of $H/M$ goes, approximately, as $-[J^{\prime
}Ms(M)]/M^{2}=-J^{\prime }s(M)/M$. When $s(M)$ becomes linear in $M$, the
carriers cease to contribute to the slope of $H/M$ $vs.$ $M^{2}$, which
becomes positive.

The change in the number of carriers with applied field, or with
magnetization is more pronounced when the initial carrier density is lowest,
as is observed in the temperature dependence of the thermopower for Tl$%
_{1.8} $Cd$_{0.2}$Mn$_{2}$O$_{7}$ in Fig. 1. This effect will contribute
strongly to the observed magnetoresistance.

We conclude that the character of the magnetic phase transition in Tl$_{2}$Mn%
$_{2}$O$_{7}$ CMR pyrochlores is first order, in contrast to previous studies
\cite{Zhao}. This first order character is compatible with a complex
ordering mechanism composed of a direct superexchange interaction between Mn
ions, and a local Kondo-type indirect coupling between Mn ions mediated by
the low density of conduction electrons. The ordering temperature seems to
be diretly related to this density of conduction electrons (i.e. the higher T%
$_{C}$, the higher density of carriers, as is the case for Tl$_{2}$Mn$_{1.8}$%
Sb$_{0.2}$O$_{7}$). Moreover, the CMR effect seems not to be related to the
first order character of the magnetic transition, since CMR is rather weak
for Tl$_{2}$Mn$_{1.8}$Sb$_{0.2}$O$_{7}$, which presents a clear first order
character. The higher CMR is obtained when the charge carriers density is
very low, and it seems to vary strongly around T$_{C}$ (i.e. the case of Tl$%
_{1.8}$Cd$_{0.2}$Mn$_{2}$O$_{7}$). All these elements (first order
character, variation of CMR and density of carriers) are obtained from a
simple Hamiltonian solved in a mean field approximation\cite{GGA00}, which
reproduces all of the above mentioned effects and also predicts a phase
separation.

We thank the financial support of MCyT to the project MAT99-1045 and
MAT2001-0539.

\widetext

\begin{figure}[tbp]
\resizebox{10cm}{!}{\rotatebox{90}{\includegraphics{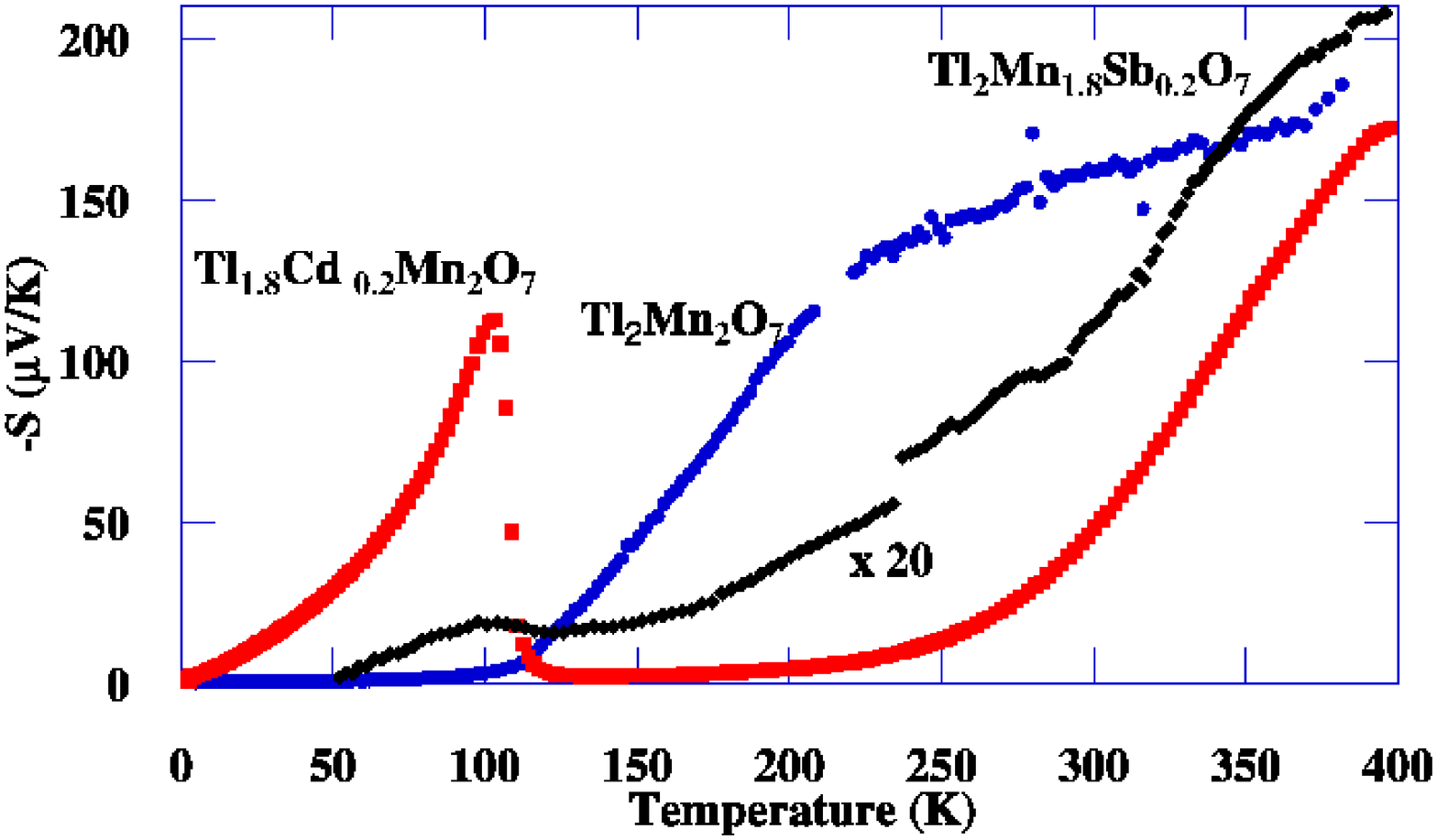}}}
\caption{Temperature dependence of the thermopower for Tl$_{2}$Mn$_{2}$O$%
_{7} $, Tl$_{1.8}$Cd$_{0.2}$Mn$_{2}$O$_{7} $ and Tl$_{2}$Mn$_{1.8}$Sb$_{0.2} 
$O$_{7}$.}
\end{figure}

\begin{figure}[tbp]
\resizebox{10cm}{!}{\rotatebox{90}{\includegraphics{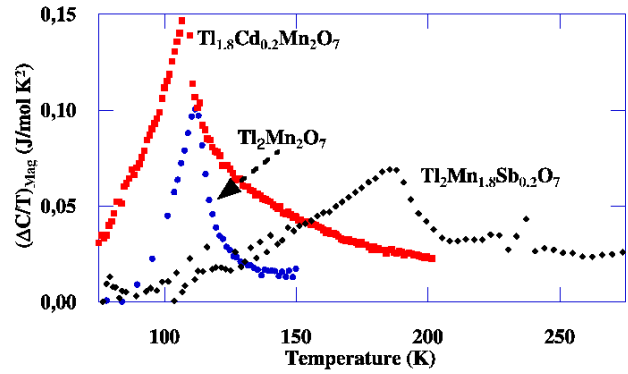}}}
\caption{Temperature dependence of the magnetic and electronic specific heat
divided by T for Tl$_{2}$Mn$_{2}$O$_{7}$, Tl$_{1.8}$ Cd$_{0.2}$Mn$_{2}$O$%
_{7} $ and Tl$_{2}$Mn$_{1.8}$Sb$_{0.2}$O$_{7}$.}
\end{figure}

\begin{figure}[tbp]
\resizebox{10cm}{!}{\rotatebox{90}{\includegraphics{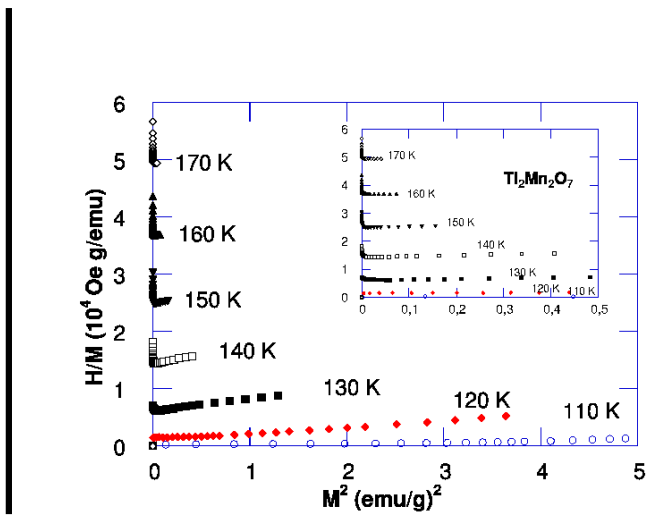}}}
\caption{H/M $vs.$ M$^{2}$ plot of the magnetization $vs.$ magnetic field
isotherms of Tl$_2$Mn$_2$O$_7$ around T$_C$. Note the onset, at low fields,
of a negative slope at a temperature near the critical point. Inset: Detail
for higher temperatures.}
\end{figure}

\begin{figure}[tbp]
\resizebox{10cm}{!}{\rotatebox{90}{\includegraphics{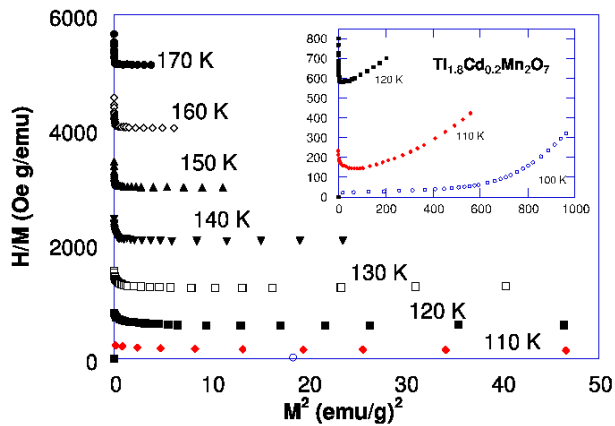}}}
\caption{ Detail of the H/M $vs.$ M$^2$ isotherms of Tl$_{1.8}$Cd$_{0.2}$Mn$%
_2$O$_7$. As in the undoped system, the negative slope starts near T$_C$.
Inset: H/M $vs.$ M$^2$ isotherms around T$_C$.}
\end{figure}

\begin{figure}[tbp]
\resizebox{10cm}{!}{\rotatebox{90}{\includegraphics{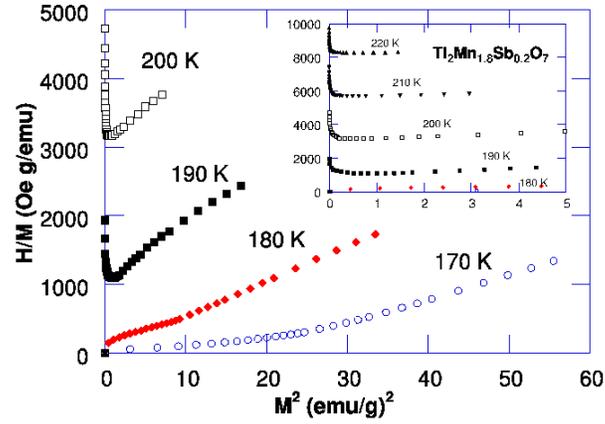}}}
\caption{ H/M $vs.$ M$^2$ isotherms, around T$_C$, of Tl$_2$Mn$_{1.8}$Sb$%
_{0.2}$O$_7$. The negative slope starts at about 190 K. Inset: Detail of
some isotherms at higher temperatures.}
\end{figure}

\begin{figure}[tbp]
\resizebox{10cm}{!}{\rotatebox{-0}{\includegraphics{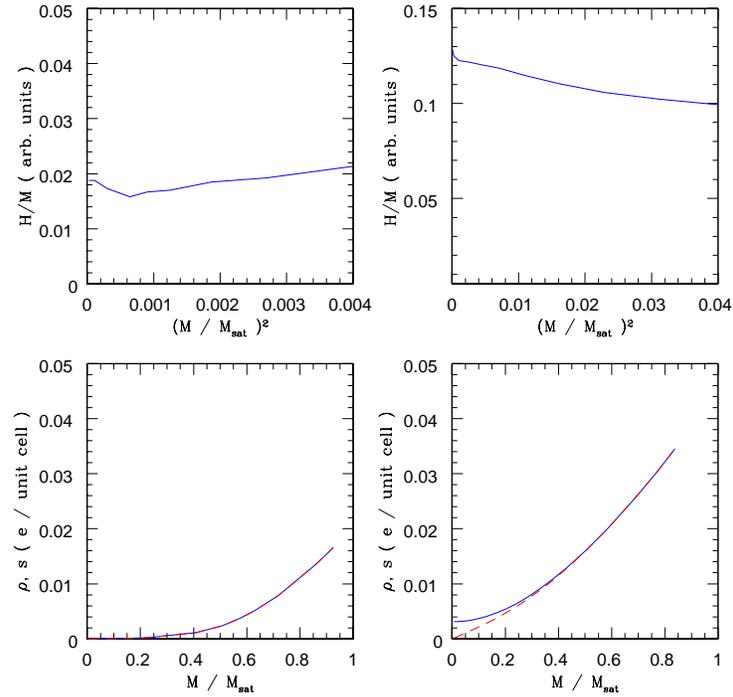}}}
\caption{ Top: calculated values of $H / M$ $vs.$ $M^2$ for low density of
carriers (left) and for higher density of carriers (right). Bottom: Carrier
density (full line) and polarization (broken line) as function of the
magnetization of the Mn ions for the same values of the chemical potential
used in the top figures (see text for details).}
\end{figure}
\newpage

\end{document}